\documentclass[preprint2]{aastex62}

\usepackage{amsmath}
\usepackage{amssymb}
\usepackage{graphicx}% Include figure files
\usepackage{dcolumn}% Align table columns on decimal point
\usepackage{bm}% bold math
\usepackage{color}

\begin{document}

\title{Evolution of fission-ignited supernova properties with uranium enrichment}

\author{Alex Deibel}\email{adeibel@iu.edu}
\affiliation{Department of Astronomy, Indiana University,
                  Bloomington, IN 47405, USA}

\author{C. J. Horowitz}\email{horowit@indiana.edu}
\affiliation{Center for Exploration of Energy and Matter and
                  Department of Physics, Indiana University,
                  Bloomington, IN 47405, USA}

\author{M. E. Caplan}
 \email{mecapl1@ilstu.edu}
\affiliation{
 Department of Physics, Illinois State University, Normal, IL 61790, USA %\\
}%           

%[0000-0002-2531-8854]                                    

%\date{\today}

\begin{abstract}
Type\textrm{--}Ia supernovae (SN Ia) are powerful stellar explosions that provide important distance indicators in cosmology. There is significant tension between values of the Hubble constant (expansion rate of the universe) determined from SN Ia and from other data. Recently, we proposed a new SN Ia mechanism that involves a nuclear fission chain reaction in an isolated white dwarf [PRL {\bf 126}, 1311010].  We find the average mass of an exploding star decreases with increasing enrichment $f_5$\textrm{---}the fraction of uranium that is the isotope $^{235}$U.  As a result, the average SN Ia luminosity decreases with increasing $f_5$.   
Furthermore, $f_5$ is likely higher in the host galaxies of SN Ia observed at large redshift $z$ because of younger galaxy ages. This change of $f_5$ leads to the evolution of SN Ia properties with redshift. If $f_5$ increases with redshift this results in an increased SN Ia rate, but a lower average SN Ia luminosity.  
\end{abstract}

\section{Introduction}
Type\textrm{--}Ia supernovae (SN Ia) are great stellar explosions that provide important distance indicators in cosmology \cite{Abbott_2019,SN_cosmology,Sullivan2010}.   They can be observed at great distances and appear to have a standardizable luminosity that can be inferred from other observations \cite{1996ApJ...473...88R,Phillips_1999,Goldhaber_2001,Phillips2017,Hayden_2019}.  This allows a precise determination of the expansion rate of the universe known as the Hubble constant.   There is significant tension, however, between the Hubble constant determined from SN Ia and the value determined from other data \cite{Riess:2016,Riess_2021,di_valentino_2021}.   Given this tension, it is important to understand possible changes in SN Ia properties that may lead to systematic error.          

Traditionally, SN Ia are thought to involve the thermonuclear explosion of a C/O white dwarf (WD) in a binary system. Here the companion is either a conventional star (single-degenerate mechanism) or another WD (double-degenerate) \cite{2012NewAR..56..122W,hillebrandt2013understanding,RUIZLAPUENTE201415}. Unfortunately, we do not yet have a complete theoretical understanding of the SN Ia mechanism or the demographics of progenitor systems. This lack of understanding makes it difficult to theoretically rule out unexpected sources of systematic errors that could lead to bias in the extraction of precision cosmological measurements. 
Changes in observed SN Ia properties may also bias the extraction of cosmological parameters. Observations find that SN Ia standardized magnitudes \cite{Hayden_2019} depend on host galaxy mass \cite{1996AJ....112.2391H,Kelly_2010} or possibly star formation activity \cite{Rigault_2015,2020A&A...644A.176R} or progenitor age \cite{lee2021discovery}.

These SN Ia properties might be explained with a newly proposed SN Ia mechanism that involves a nuclear fission chain-reaction igniting carbon burning in an {\it isolated} WD \cite{PhysRevLett.126.131101}.  The model is further explored in the subsequent papers \cite{fission2,fission_network:2021}. 
The first solids that crystallize in a cooling WD are expected to be uranium-rich because of uranium's high atomic charge $Z=92$\textrm{---} forming a crystal with a ratio $R_U$ of $^{235}$U to $^{238}$U isotopes.    
The enrichment $f_5$ is the fraction of all uranium that is $^{235}$U and is the natural quantity to use in calculations of fission chain reactions and criticality. The enrichment and ratio are related by $f_5=R_U/(1+R_U)$. If $f_5$ is high enough, roughly $\ge 0.12$, criticality is reached and a fission chain-reaction in the uranium crystal burns available fissile fuel. The chain-reaction and breeder reaction release $\approx 0.36 \, \mathrm{MeV/nucleon}$ and raise the final temperature to $T_f \approx 6 \times 10^9 \, \mathrm{K}$~\citep{fission_network:2021}. Hydrodynamical simulations need to be performed to investigate if the heat release triggers carbon fusion or possibly a detonation of the WD. 

Within the fission mechanism the dependence on enrichment naturally leads to the evolution of SN Ia properties with redshift. This is because the enrichment is expected to change with time: decreasing as $^{235}$U decays with half-life 700 Myr and increasing as new uranium is synthesized in $r$-process events.  Furthermore, $f_5$ may differ from one galaxy or location to another depending on its nucleosynthesis history. This means that the ratio $R_U$ changes with time and can be used as a cosmo-chronometer \cite{Truran1998}. The fission mechanism, because it depends on $f_5$ means {\it SN Ia effectively radioactively date their host galaxies}.  SN Ia at large redshift come from younger galaxies (in general) than nearby SN Ia. Within the fission model, {\it SN Ia at high redshift have different properties than nearby ones}. In particular, larger $f_5$ values may allow lower mass WDs to explode and SN Ia global properties (such as rate and average luminosity) will therefore evolve with redshift.

In this letter we make a simple model of galactic chemical evolution and how $f_5$ may change with galactic age.  We then review the fission mechanism and discuss how $f_5$ determines which stars explode.  We find that lower mass stars may explode for increasing $f_5$.  This may lead to SN Ia luminosities that decrease with increasing $f_5$. We end with a discussion of the evolution of $f_5$ with redshift and galactic age.

 \section{Uranium enrichment and galactic chemical evolution} 
 We start with a simple model for the enrichment of the Milky Way (in the neighborhood of the solar system) vs galactic age $t$.  We assume the galaxy formed about $1\, \mathrm{Gyr}$ after the Big Bang so $t$ is the age of the universe minus $1 \, \mathrm{Gyr}$.  Our first model has $r$-process nucleosynthesis events \cite{Horowitz_2019,physicsreports_rprocess,2001AA...379.1113G}, such as neutron star mergers, occurring at a constant rate and assumes each $r$-process event produces uranium with the same initial ratio $R_U^0$.  We then integrate all uranium production up to a final time $t_f$ and include radioactive decay during the time between production and $t_f$.  This gives,
\begin{equation}
R_U(t_f)=R_U^0\frac{\int_0^{t_f} {\rm exp}{(-\frac{t_f-t}{\tau_5})}dt 
}{\int_0^{t_f}{\rm exp}{(-\frac{t_f-t}{\tau_8})}dt 
}\, .
\label{eq.Ru}
\end{equation}

Here $\tau_5= (0.7/\ln 2) \, \mathrm{Gyr}$ is the $^{235}$U mean life and $\tau_8=(4.5/\ln 2) \, \mathrm{Gyr}$ is the mean life of $^{238}$U. Note that if a new star is formed at $t_f$ it will have an enrichment  $f_5(t_f)=R_U(t_f)/(1+R_U(t_f))$. 

\begin{figure}[tb]
\centering  %left bottom top right)
\includegraphics[width=0.48\textwidth]{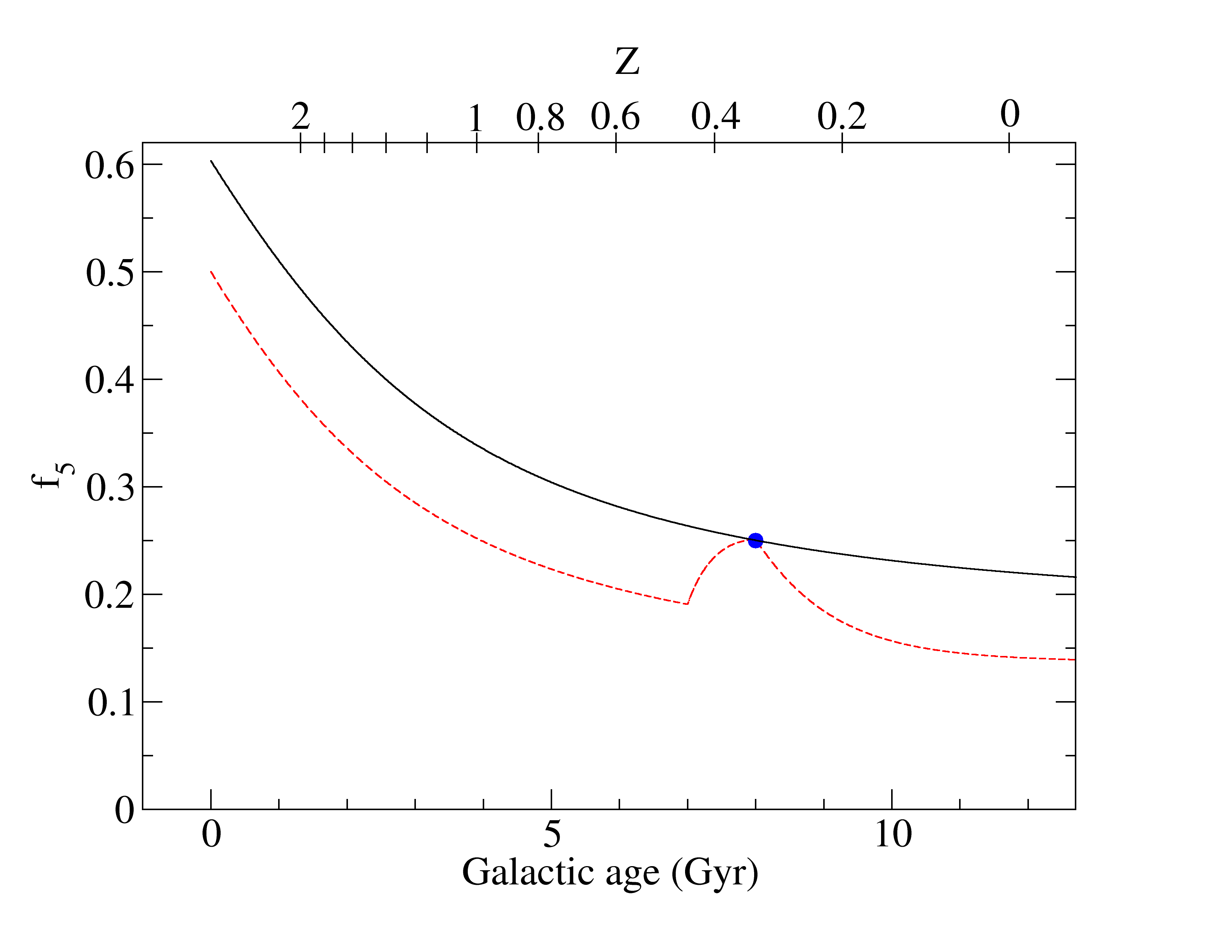}
\caption{\label{Fig3} Uranium enrichment $f_5$ versus galactic age assuming a constant $r$-process production rate (solid-black curve) or a constant rate augmented by a higher rate for $1\, \mathrm{Gyr}$ before solar system formation (dashed-red curve).  The filled circle shows the enrichment of the solar system at formation. The top axis shows the age (at star formation) of a galaxy hosting a SN Ia that is observed at redshift $z$.  This assumes a delay time of $\approx 1 \, \mathrm{Gyr}$ from star formation until SN Ia.}	
\end{figure}

To fit $R_U^0$ we must know $f_5$ at some point in Galactic history. Presently, the only measurement of $f_5$ is from the solar system, $f_5=0.007$ observed today \cite{lodders2019solar}, which gives an enrichment of the solar system at the time of formation of $f_5^{ss}\approx 0.25$. Assuming a constant $r$-process rate we obtain $R_U^0\approx 1.52$; this value is slightly larger than $R_U^0\approx 1$ suggested by $r$-process nucleosynthesis simulations \cite{physicsreports_rprocess}\footnote{Erika Holmbeck private communication}.  Therefore, for our second model we fix $R_U^0=1$ and assume that $r$-process nucleosynthesis occurred at a higher rate for the $1 \, \mathrm{Gyr}$ before solar system formation.  We then fit this higher rate to reproduce $f_5^{ss}$ and this gives the dashed line in Fig. \ref{Fig3}.

Figure \ref{Fig3} shows that the enrichment $f_5(t)$ is larger at earlier times.  When the galaxy was young there was little time to accumulate additional $^{238}$U and $f_5$ is close to the large $r$-process initial value.  Indeed at $t=0$ one has $f_5(0)=R_U^0/(1+R_U^0) \gtrsim 0.5$.  As the galaxy ages, $^{238}$U accumulates from $r$-process events that occurred long enough in the past that the produced $^{235}$U has decayed.  As a result, $f_5(t)$ decreases with increasing $t$.

\begin{figure}[tb]
\centering  %left bottom top right)
\includegraphics[width=0.48\textwidth]{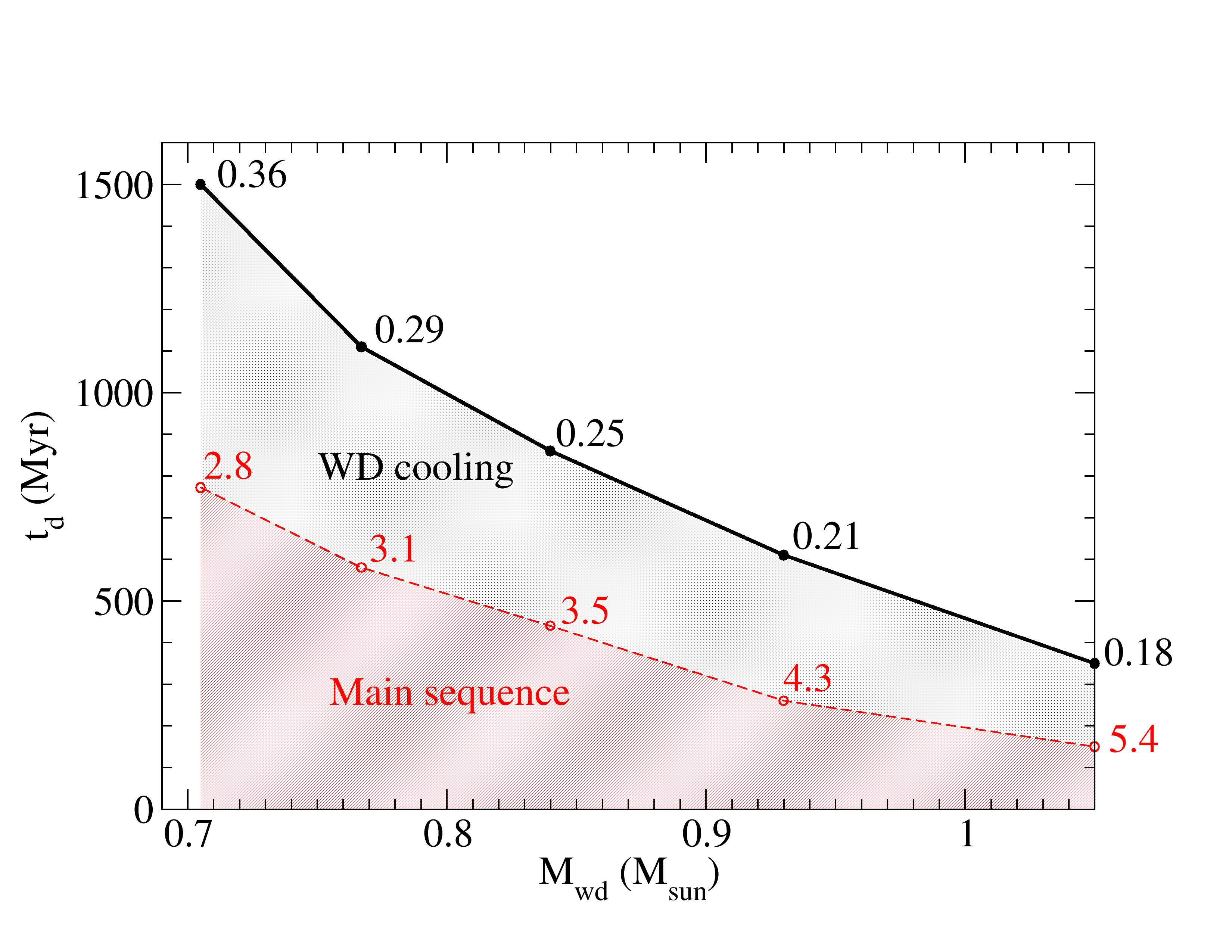}
\caption{\label{Fig1} Delay time $t_d$ from star formation until U crystallization (solid black line) versus WD mass.   The black numbers, next to the solid line, are the minimum $f_5(t_f)$ necessary at star formation to have $f_5$ of at least 0.14 when U crystallizes.  The red shaded region shows the main sequence lifetime and the red numbers indicate the original main sequence mass in M$_\odot$  \cite{Cummings_2018}.  The black shaded region shows the WD cooling time until U crystallization. }	
\end{figure}

\section{The fission SN Ia scenario} 
The fission mechanism for SN Ia depends crucially on $f_5$.  A pre-stellar cloud with an initial enrichment $f_5$ collapses to form a star, that after its main sequence lifetime evolves, typically with mass loss, to form a WD  \cite{Cummings_2018}.  The WD then cools until the very first uranium-rich solids form and a fission chain reaction becomes possible. The uranium crystallizes at about twice the temperature that carbon and oxygen crystallize  \cite{PhysRevLett.126.131101}.

These steps involve a total delay time $t_d$ until a SN Ia that depends on WD mass as shown in Fig. \ref{Fig1}. During this delay time $f_5$ decreases as $^{235}$U decays. If $f_5(t_f)$ is the initial enrichment at star formation, the enrichment at time $t$ is 
 %\begin{equation}
 $f_5(t)=\{(f_5(t_f)^{-1}-1)\exp[(t-t_f)/\tau]+1\}^{-1}$ 
% \label{eq.ft}
% \end{equation}
 with $\tau=1.2\, \mathrm{Gyr}$.  If $f_5(t_d+t_f)$ is too small carbon ignition may not be possible.  Ref. \cite{PhysRevLett.126.131101} found that $f_5>0.12$ for the system to be critical.  The enrichment may need to be slightly larger than this in order to have a robust chain reaction that can lead to carbon ignition.  We therefore assume that $f_5(t_d+t_f)> 0.14$ is necessary for the star to explode via the fission mechanism.  

The $f_5(t_f)$ values that are necessary at star formation in order to have $f_5(t_d+t_f)=0.14$ after the delay time are also listed in Fig. \ref{Fig1}.  The delay time increases rapidly for a lower mass WD both because of a longer main sequence lifetime and because of a longer WD cooling time.  Therefore, the necessary $f_5$ increases with decreasing WD mass.  

The maximum WD mass that will explode may be set by its composition. We assume a WD above about 1.05 M$_\odot$ has an O/Ne core and will not explode because fission may not sufficiently raise the temperature for O burning.  Therefore, WDs between a minimum mass $M_{\mathrm{min}}$ determined by $f_5$ and 1.05 M$_\odot$ may explode via fission.  From Fig. \ref{Fig1} we have the following fit for $M_{\mathrm{min}}(f_5)$.  For $f_5<0.18$ no stars explode.  For $0.18\leq f_5 \leq 0.5$,
\begin{equation}
M_{\mathrm{min}}\approx (0.034 + 0.600 f_5 + 0.163/f_5)\  {\rm M}_\odot\, .
\end{equation}
For $f_5>0.5$ the delay time increases very rapidly for $M_{\mathrm{min}}$ and we therefore set $M_{\mathrm{min}} = 0.660\,\mathrm{M}_\odot$ .  The mass range of WDs that may explode via fission are illustrated in Fig. \ref{Fig2}.

\section{Supernova luminosity and evolution with enrichment} 
We now discuss predictions for the absolute luminosity of nearby SN Ia. Within the fission scenario, a nuclear chain-reaction produces a deflagration near the center of the WD that may turn into a detonation later (for example, \cite{Poludnenkoeaau7365}). Indeed, reaction network simulations of the fission chain-reaction show a final temperature above $T = 5 \times 10^9 \, \mathrm{K}$ \citep{fission_network:2021}. There are, however, no previous simulations of carbon ignition (but see \citet{1992ApJ...396..649T}), deflagrations in a sub-Chandrasekhar mass WD, or the possibility of a deflagration-to-detonation transition (DDT).  If a DDT happens quickly, the result may be similar to an initial detonation and we therefore consider detonation simulations to illustrate one possibility.   

\begin{figure}
\centering  %left bottom top right)
\includegraphics[width=0.48\textwidth]{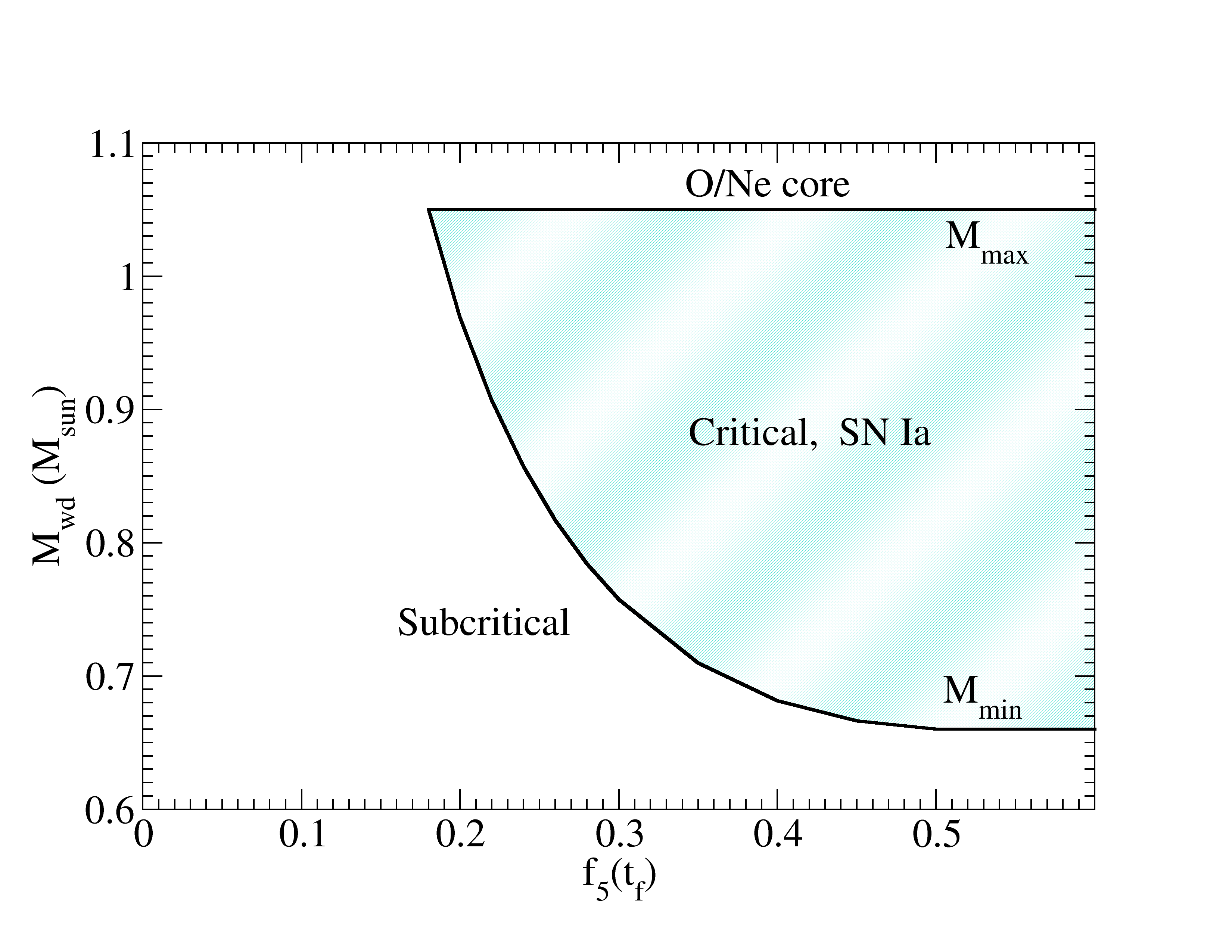}
\caption{\label{Fig2} The colored region indicates the WD masses that may explode via fission ignition as a function of the uranium enrichment at the time of star formation $f_5(t_f)$.  
}	
\end{figure}

\begin{figure}[tb]
\centering  %left bottom top right)
\includegraphics[width=0.48\textwidth]{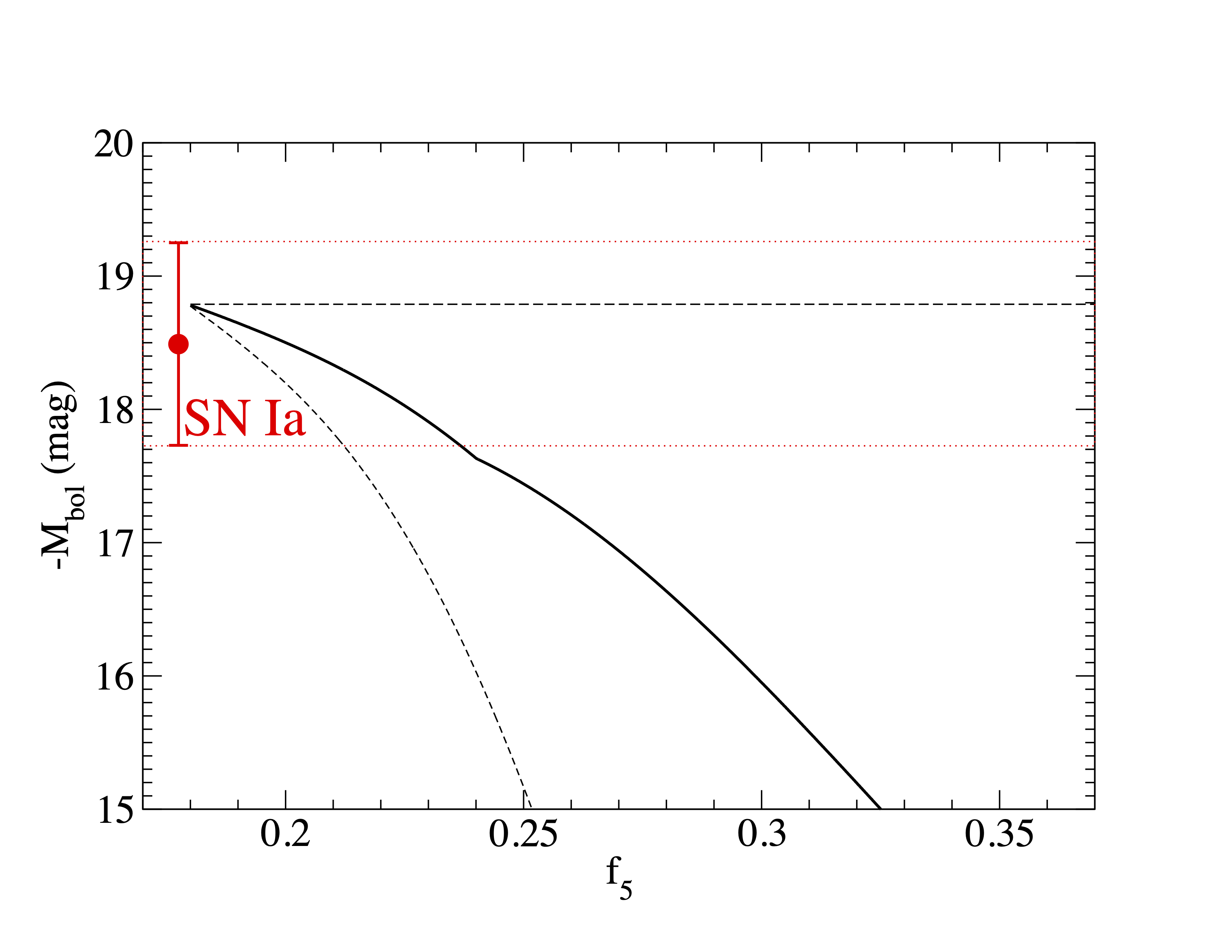}
\caption{\label{Fig4} Maximum in the SN Ia bolometric light curve versus uranium enrichment $f_5$. The red point and error bar shows SN Ia luminosity observations \cite{10.1111/j.1365-2966.2011.18160.x}.  The dashed black lines show one dim. hydrodynamical simulation results \cite{10.1093/mnras/sts423} for the maximum (top) and minimum (bottom) WD masses from Fig. \ref{Fig2} and the solid black line shows the average luminosity (see text). }
\end{figure}

One dimensional hydrodynamical simulations of a pure detonation scenario place a detonation in the center of a C/O WD of mass $M_{\rm wd}$ and the maximum in the bolometric light curve is then determined \cite{2010ApJ...714L..52S,10.1093/mnras/sts423}.  We assume these simulations roughly apply for the fission scenario and we use a polynomial fit  to these simulations \cite{10.1093/mnras/sts423} that gives luminosity as a function of WD mass $L(M_{\rm wd})$.  We plot $L$ in Fig. \ref{Fig4} for the maximum $M_{\mathrm{max}}$ and minimum mass $M_{\mathrm{min}}$ that will explode for a given $f_5$ as shown in Fig. \ref{Fig2}. We also show the average luminosity where WD masses between $M_{\rm min}$ and $M_{\rm max}$ are weighted by a main sequence initial mass function $\propto M_{\rm ms}^{-2.35}$.

Because the $M_{\rm min}$ that can explode depends on $f_5$, the SN Ia luminosity will also evolve with enrichment. Fig. \ref{Fig2} shows that as $f_5$ increases smaller WD masses can explode after somewhat longer delay times.  The enrichment-dependent bolometric magnitude as shown in Fig. \ref{Fig4} is approximately,
\begin{equation}
    M_{\rm bol}\approx -36.00 + 50.53\, f_5 + 1.477/f_5\ \, {\rm mag} \ ,
\end{equation}
for $0.18\le f_5\le 0.34$.  For larger $f_5$ see \footnote{For $0.34<f_5\le 0.5$ we have $M_{\rm bol}\approx 59.59-57.414f_5-18.58/f_5$ and for $f_5>0.5$ we have $M_{\rm bol}\approx -6.33$}.  This means that the SN Ia luminosity decreases as $f_5$ increases.  Therefore, {\it SN Ia properties will evolve with $f_5$.}   As can be seen in Fig. \ref{Fig4}, for $0.18<f_5\lessapprox 0.25$, the predicted luminosities are in good agreement  with the average luminosity of nearby SN Ia observed in the Lick Observatory SN search %$M_{\rm bol}=-18.49 \pm 0.76 \, \mathrm{mag}$ 
\cite{10.1111/j.1365-2966.2011.18160.x}.

\begin{figure}[tb]
\centering  %left bottom top right)
\includegraphics[width=0.48\textwidth]{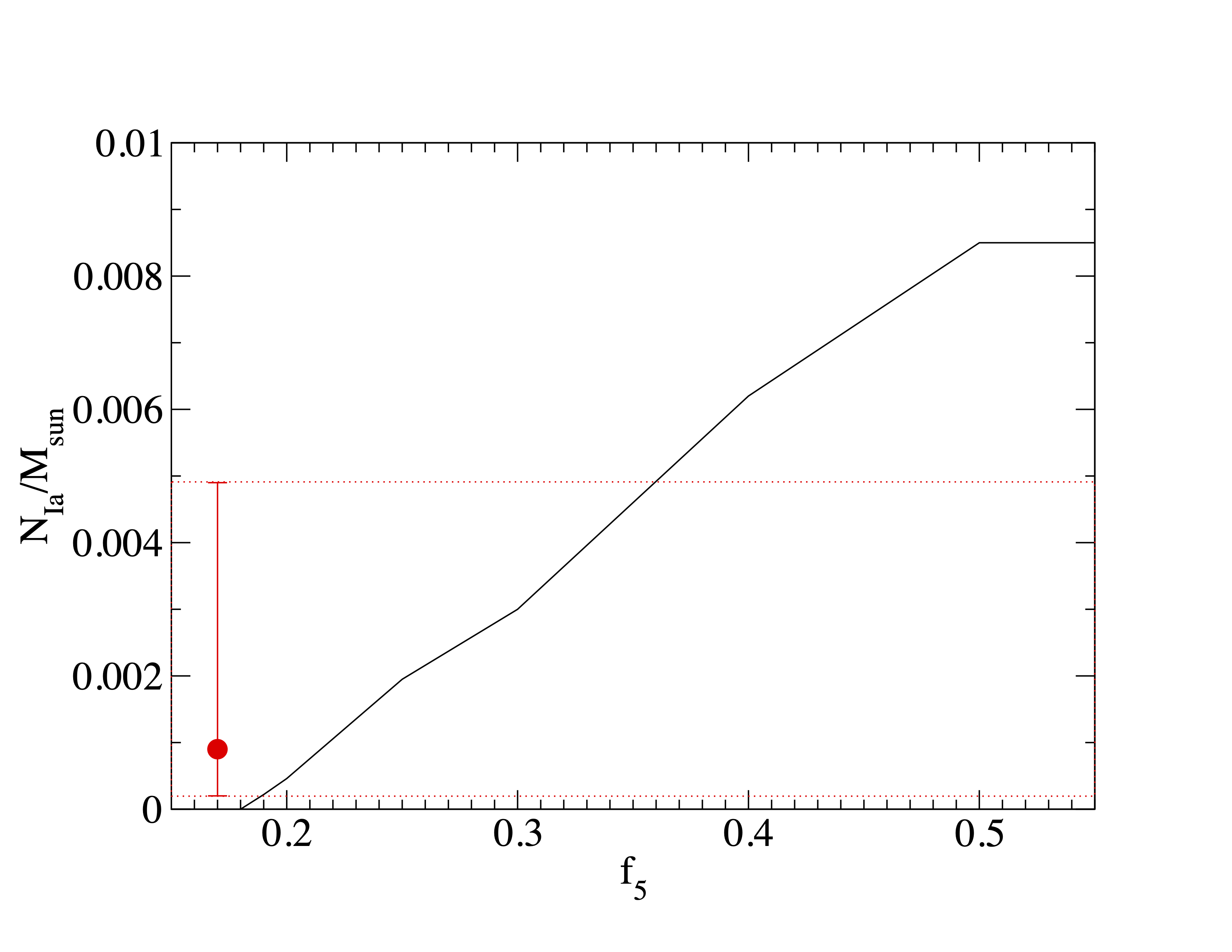}
\caption{\label{Fig4b} Type Ia SN production efficiency, in SN per solar mass of stars, versus uranium enrichment $f_5$. The black curve shows the calculated, and the red point and error bar the observed \cite{10.1093/mnras/stab1943} production efficiency. 
}
\end{figure}

The SN Ia production efficiency is the number of SNe that eventually occur per $M_\odot$ of stars.  We calculate this as the number of main sequence stars that produce WD with masses between $M_{\rm min}$ and $M_{\rm max}$ divided by the total mass in stars (for an initial mass function $\propto M_{ms}^{-2.35}$),
\begin{equation}
    \frac{{\rm N}_{Ia}}{M_*}\approx 0.0263(f_5-0.18)\ \ {\rm SNe\, }M_\odot^{-1},
\end{equation}
for $0.18<f_5<0.5$ and N$_{Ia}/M_*=0.0084$ SNe $M^{-1}_\odot$ for $f_5>0.5$.  This is plotted in Fig. \ref{Fig4b} and agrees with Dark Energy Survey observations \cite{10.1093/mnras/stab1943} for $0.19\le f_5\le 0.36$.  The SN Ia production efficiency increases with $f_5$ as a larger range of WD masses explode.

\begin{figure}[tb]
\centering  %left bottom top right)
\includegraphics[width=0.48\textwidth]{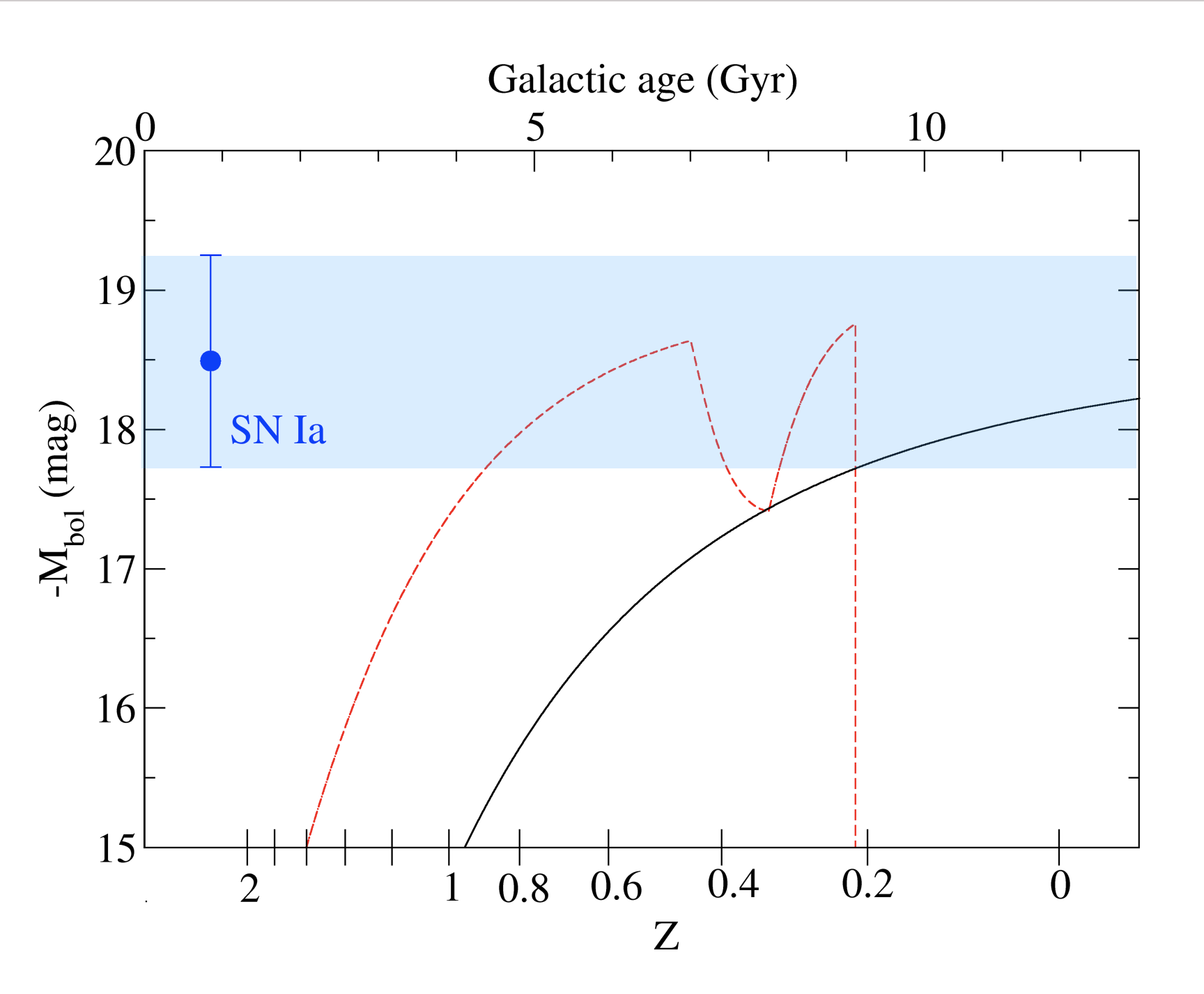}
\caption{\label{Fig5} Average SN Ia luminosity vs redshift $z$.  The average luminosity from Fig. \ref{Fig4} is plotted for different models of uranium enrichment. The solid curve and the dashed curve correspond to the solid and dashed curves in Fig. \ref{Fig3}. The average of SN Ia observations is shown as a blue error band \cite{10.1111/j.1365-2966.2011.18160.x}.  }
\end{figure}

\begin{figure}[tb]
\centering  %left bottom top right)
\includegraphics[width=0.48\textwidth]{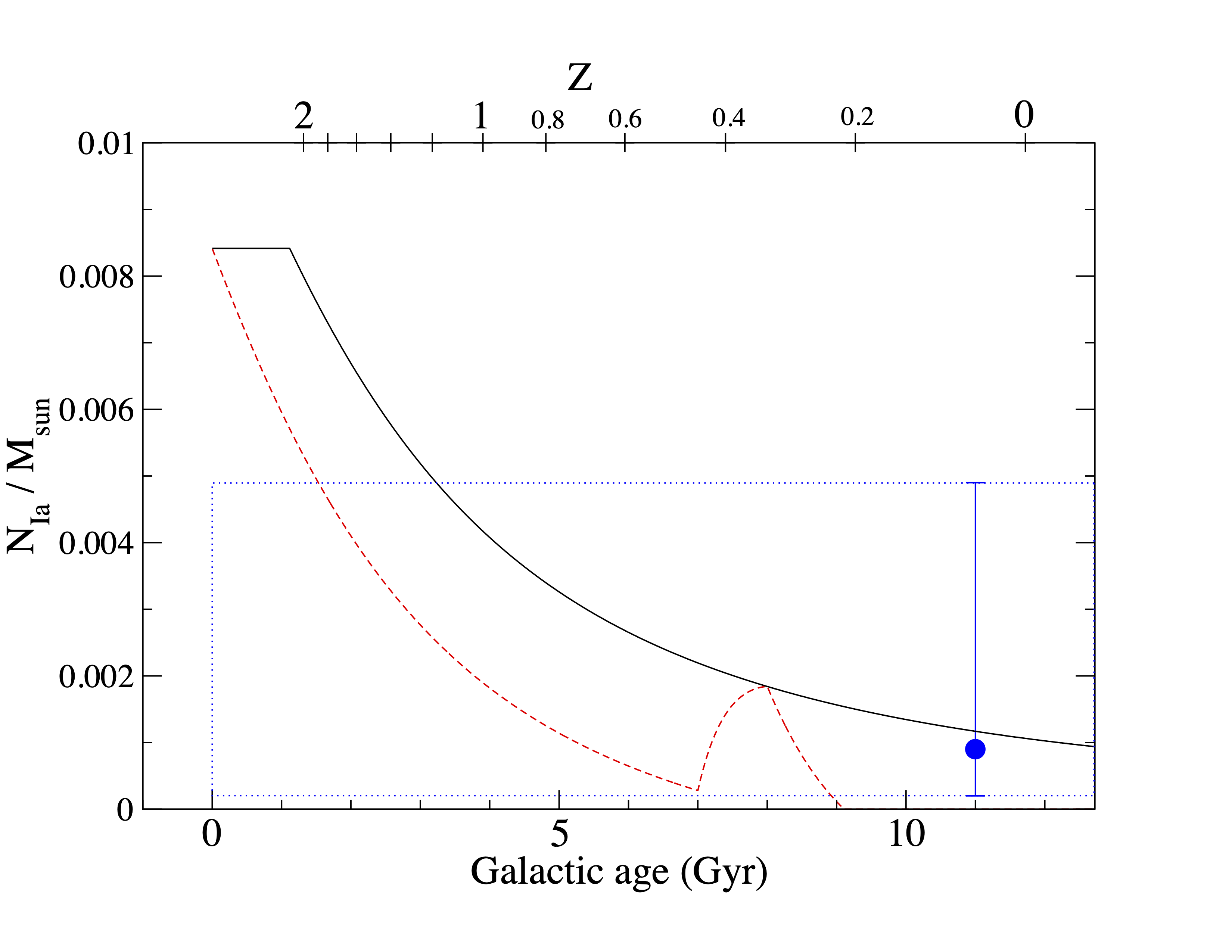}
\caption{\label{Fig7} Production efficiency in SN per solar mass versus galactic age (lower x axis) or redshift (upper x axis).  The black solid and red dashed curves use the corresponding models of uranium enrichment from Fig. \ref{Fig3}.  The observed production efficiency from the Dark Energy Survey is shown as a blue data point \cite{10.1093/mnras/stab1943}.  }
\end{figure}

 The evolution of $f_5$ with redshift (Fig. \ref{Fig3}) will produce changes in SN Ia luminosity with redshift, as shown in Fig. \ref{Fig5}. As $f_5$ increases with increasing $z$, this allows WDs of lower mass to explode and the average luminosity decreases (in general) with increasing $z$. We show in Fig. \ref{Fig7} the SN Ia rate as it follows the $f_5$ evolution with redshift. The SN Ia rate should increase with increasing redshift because a greater number of WDs will explode at higher $f_5$ values (see Fig. \ref{Fig2}).  
 
 \section{Uranium elemental and enrichment observations} 
 The enrichment in Fig. \ref{Fig3}, or from more sophisticated galactic chemical evolution calculations, may be model dependent.    Unfortunately, we only know $f_5$ for the solar system.  Additional observations of $f_5$ in the present Galaxy would be a useful constraint.  Figure \ref{Fig3} predicts that $f_5$ in any star that is young (compared to the 700 Myr $^{235}$U half-life) will be large, over 10\%.  This is much larger than the present 0.7\% enrichment of the solar system.  This is because a young star is made from gas that until recently was receiving fresh $r$-process material that is rich in $^{235}$U.  The elemental abundance of uranium is notoriously difficult to measure in stellar spectra.  However there are some observations in old metal poor stars \cite{U_oldstar,Frebel_2007}.   If it is possible to observe U in a young star, determining the isotopic ratio may not be so much harder.  The isotope shift is about one part in 35,000 \cite{PIETSCH1998751} and we predict the amount of $^{235}$U is relatively large.  Such an observation, if possible, would help constrain the temporal and spatial variation in $f_5$ in the Galaxy.

\section{Discussion} 
The fission model for SN Ia depends crucially on the enrichment $f_5$\textrm{---}the fraction of all uranium that is the isotope $^{235}$U. The enrichment decreases with time as $^{235}$U decays and $^{238}$U accumulates.

We expect multiple episodes of star formation and the mixing of different gas volumes with different enrichments.  This may have an averaging effect and lead to an $f_5$ closer to an average value with smaller amplitude changes during star formation or passive times.  Still, the enrichment, almost certainly, is not homogeneous and may vary both within a galaxy and from one galaxy to the next.  Regions of higher enrichment will have more SN Ia, see Fig.~\ref{Fig4b}.  

As we show in Fig. \ref{Fig2}, a larger $f_5$ will allow lower mass WDs to explode after a somewhat longer delay time. SN Ia from lower mass WDs will have lower luminosities. Thus the SN Ia average luminosity decreases with increasing $f_5$, as shown in Fig. \ref{Fig4}. Because $f_5$ is larger in young galaxies the $f_5$ will also increase with redshift; resulting in a lower average SN Ia luminosity at higher redshift.
 
These changes in SN Ia properties with redshift could be important for precise measurements of cosmological parameters \textrm{--} like the Hubble constant \textrm{--} if phenomenological corrections used to standardize SN Ia luminosities are imperfect.   To better understand these and other possible systematic errors it is important to better understand the SN Ia explosion mechanism and the progenitor systems. 

The possibility of systematic error may be particularly clear in one limiting case.  The dashed curves in Figs. \ref{Fig3} and \ref{Fig5} have $f_5<0.18$ for $z<0.2$.  Within this model for $f_5$, there will be no fission ignited SN Ia to the right of the vertical dashed line in Fig. \ref{Fig5}.  However, $f_5$ is larger for larger $z$ so that there are fission ignited SN Ia that can only be observed at high $z$.  Therefore, no matter how one trains corrections to the luminosities of local SN, one would not necessarily account for and calibrate the fission-ignited SN. Perhaps local SN and fission-ignited SN would have different Phillips relations. As a result, one may incorrectly deduce their distance.  This case would be analogous to the Oklo natural nuclear reactor which only operated $\approx 2\, \mathrm{Gyr}$ ago when $f_5$ was larger \cite{GAUTHIERLAFAYE19964831,PhysRevLett.93.182302,10.2307/24950391}.  Such a natural reactor is likely impossible today because $f_5$ is too small. 

  Perhaps surprisingly, the fission model does not seem to depend strongly on metallicity.  A lower metallicity, including lower U elemental abundance, leads to U crystallizing at a slightly lower temperature (but still significantly hotter than the C/O melting temperature).  This could lead to a slightly longer delay time.  However, the crystal when it does form is still expected to be very U rich and will still be critical for appropriate $f_5$ values. Therefore, fission ignition may work for a broad range of metallicities.  

Note that in this manuscript we have discussed the possibility of a fission chain-reaction igniting a pure detonation of an isolated WD. This is one possible outcome of the heat deposition from nuclear burning in the U-rich crystal. Hydrodynamical simulations of carbon burning in the WD core are necessary to explore the parameter phase-space and support this as a possibility. It may be that a carbon deflagration occurs without a DDT or the DDT occurs very late, and isolated WDs instead produce subluminous SN.    

\section{Conclusion} 
The fission model for SN Ia depends crucially on the enrichment $f_5$.  This is the fraction of all uranium that is the isotope $^{235}$U. The enrichment evolves with time as U decays and fresh $^{235}$U is produced in $r$-process nucleosynthesis events. Crucially, since $^{235}$U and $^{238}$U have different half-lives but are produced in nearly equal amounts in $r$-process events, $f_5$ is decreasing with cosmic time.   A larger $f_5$ will allow lower mass WD to explode after a somewhat longer delay time and these may produce SN Ia with lower luminosities.  Thus the SN Ia average luminosity decreases with increasing $f_5$. We expect $f_5$ to be larger for SN Ia observed at high redshift because of the younger age of their host galaxies. This evolution of $f_5$ with redshift will lead to changes in SN Ia luminosity or other properties that are potentially important for precision cosmological measurements.

{\it Acknowledgements:} We thank Erika Holmbeck for many helpful suggestions and for sharing preliminary results on uranium synthesis in the $r$-process.   We thank Catherine Pilachowski for helpful suggestions on uranium observations.  We thank Ezra Booker, Constantine Deliyannis, Tomasz Plewa, and Rebecca Surman for helpful discussions.  The work of CJH was performed in part at the Aspen Center for Physics, which is supported by National Science Foundation grant PHY-1607611. This research was supported in part by the US Department of Energy Office of Science Office of Nuclear Physics grants DE-FG02-87ER40365 and DE-SC0018083 (NUCLEI SCIDAC).

\bibliography{main}
\bibliographystyle{yahapj}

\end{document}